\documentclass{article}
\newcommand{\be}{\begin{equation}}
\newcommand{\ee}{\end{equation}}
\newcommand{\bea}{\begin{eqnarray}}
\newcommand{\eea}{\end{eqnarray}}
\newcommand{\pa}{\partial}
\newcommand{\ve}{\varepsilon}
\newcommand{\ds}{\displaystyle}

\begin{document}

\title{The binary black-hole problem at the third post-Newtonian approximation
in the orbital motion: Static part}

\author{Piotr Jaranowski
\\{\it Institute of Physics, Bia{\l}ystok University}\\
  {\it Lipowa 41, 15-424 Bia{\l}ystok, Poland}
\thanks{E-mail: pio@alpha.uwb.edu.pl}\vspace{2ex}\\
Gerhard Sch\"afer
\\{\it Theoretisch-Physikalisches Institut, 
       Friedrich-Schiller-Universit\"at}\\
  {\it Max-Wien-Platz 1, 07743 Jena, Germany}
\thanks{E-mail: gos@tpi.uni-jena.de}}

\date{}

\maketitle

\begin{abstract}

Post-Newtonian expansions of the Brill-Lindquist and Misner-Lindquist solutions
of the time-symmetric two-black-hole initial value problem are derived. The
static Hamiltonians related to the expanded solutions, after identifying the
bare masses in both solutions, are found to differ from each other at the third
post-Newtonian approximation. By shifting the position variables of the black
holes the post-Newtonian expansions of the three metrics can be made to
coincide up to the fifth post-Newtonian order resulting in identical static
Hamiltonians up the third post-Newtonian approximation. The calculations shed
light on previously performed binary point-mass calculations at the third
post-Newtonian approximation.

\vspace{0.5cm}\noindent PACS number(s):
04.25.Nx, 04.30.Db, 97.60.Jd, 97.60.Lf
\end{abstract}

\section{Introduction and Summary}

In a recent paper by the authors \cite{JS98} the claim was put forward that at
the third post-Newtonian (3PN) approximation of general relativity  point-mass
models for binary systems have to be replaced by  black-hole models to become
unique. To confirm the claim that binary point-mass models are incomplete at
the 3PN approximation, in the present paper the Brill-Lindquist (BL)
\cite{BL63} and Misner-Lindquist (ML) \cite{M63,L63} solutions of the
time-symmetric initial value problem for binary black holes are expanded into
post-Newtonian series  and post-Newtonian Hamiltonians related to these
expansions are calculated. 

The BL and ML solutions are known to differ from each other topologically as
well as geometrically \cite{CK82,AP96}. The interesting question therefore
arises if at the 3PN order of approximation in the relative motion
differences show up. The remarkable outcome of our calculations from Sec.\ 2 is
that the two solutions have different Hamiltonians starting at the 3PN order,
but also that these Hamiltonians can be made to coincide by shifting the
centers of the black holes. Especially interesting shifts are those where the
two black holes of the ML solution obtain vanishing dipole moments as to
possibly coincide with the monopolar black hole potentials of the BL solution
to higher orders.

It is evident that at the 3PN order of approximation point-mass models in
many-body systems are no longer applicable. In Sec.\ 3 we perform static (i.e.\
with linear momenta of the bodies and transverse-traceless part of the three
metric set equal to zero) binary point-mass calculations using different
regularization methods which lead to different metric coefficients and
Hamiltonians, i.e.\ we end up with a static point-mass ambiguity. Sec.\ 4 is
devoted to some consistency calculations for the regularization procedures of
Sec.\ 3 and to the comparison with previous results. Some details of the
calculations are given in two appendices.

Although it holds that through the postulate of vanishing black-hole dipole 
moments in isotropic coordinates a unique static Hamiltonian can be obtained 
at the 3PN order of approximation, the problem of finding a similar unique
total Hamiltonian is more complicated because the point-mass ambiguity detected
in the previous paper \cite{JS98} is a dynamical one which includes the momenta
of the objects as well as the radiation degrees of freedom of the gravitational
field  (transverse-traceless part of the three metric). This ambiguity lies far
beyond any known exact or approximate solutions of the Einstein field equations
and thus, it can not be resolved in a way the BL and ML solutions for the
constraint equations allow for a clear  identification of the ambiguous static
contributions. The static ambiguity has not been mentioned in the paper
\cite{JS98}.

We use units in which $16\pi G=c=1$, where $G$ is the Newtonian gravitational
constant and $c$ the velocity of light.  We employ the following notation:
${\bf x}=\left(x^i\right)$ ($i=1,2,3$) denotes a point in the 3-dimensional
Euclidean space endowed with a standard Euclidean metric and a scalar product
(denoted by a dot).  Letters $a$ and $b$ are body labels ($a,b=1,2$), so ${\bf
x}_a$ denotes the position of the $a$th body, and $m_a$ denotes its mass
parameter. We also define  ${\bf r}_a:={\bf x}-{\bf x}_a$, $r_a:=|{\bf r}_a|$,
${\bf n}_a:={\bf r}_a/r_a$; and for $a\ne b$,  ${\bf r}_{ab} := {\bf x}_a -
{\bf x}_b$, $r_{ab}:=|{\bf r}_{ab}|$, ${\bf n}_{ab}:={\bf r}_{ab}/r_{ab}$;
$|\cdot|$ stands here for the Euclidean length of a vector.  Indices with round
brackets, like in $\phi_{(2)}$, give the order of the object in inverse powers
of the velocity of light, in this case, $1/c^2$. We abbreviate
$\delta\left({\bf x}-{\bf x}_a\right)$ by $\delta_a$.

\section{PN expanded Brill-Lindquist and Misner-Lindquist solutions}

In the static case, as defined above, the three metric can be put into 
conformally flat form [cf.\ Eq.\ (3) in \cite{JS98}]
\be
\label{metric}
g_{ij}=\left(1+\frac{1}{8}\phi\right)^4\delta_{ij}.
\ee
We use the representation of the BL and ML solutions as given in Ref.\ 
\cite{AP96}. For the BL solution the function $\phi$ from Eq.\ (\ref{metric}) 
equals [cf.\ Eq.\ (3) in \cite{AP96}]
\be
\label{BL}
\phi^{\rm BL} = 8\left(\frac{\alpha_1}{r_1}+\frac{\alpha_2}{r_2}\right).
\ee
Here the positive parameters $\alpha_1$ and $\alpha_2$ can be expressed in
terms of  the bare masses $m_1$ and $m_2$ of the black holes and the coordinate
distance  $r_{12}$ between them. The relations are given in Appendix A.
The ML solution is described by the function [cf.\ Eq.\ (4) in \cite{AP96}]
\be
\label{ML}
\phi^{\rm ML} = 8\left(\frac{a}{r_1}+\frac{b}{r_2}\right)
 + 8\sum_{n=2}^{\infty}
\left(\frac{a_n}{|{\bf x}-{\bf d}_n|}+\frac{b_n}{|{\bf x}-{\bf e}_n|} \right),
\ee
where ${\bf x}_1$ in $r_1=|{\bf x}-{\bf x}_1|$ is the position of the center of
the black hole 1 of radius $a$ and ${\bf x}_2$ in $r_2=|{\bf x}-{\bf x}_2|$ is
the position of the center of the black hole 2 of radius $b$, relative to a
given origin in the flat space; ${\bf d}_n$ ($n\ge2$) are the positions of the
image poles of black hole 1, ${\bf e}_n$ ($n\ge2$) are the positions of the
image poles of black hole 2, $a_n$ and $b_n$ ($n\ge2$) are the corresponding
weights.  For the ML solution the choice of the bare masses is not as obvious
as in the case of the BL solution \cite{AP96}. We use the definition of the
bare masses introduced by Lindquist in \cite{L63}.  Then the solution
(\ref{ML}) can be iteratively expressed in terms of the bare masses $m_1$,
$m_2$ and the vector ${\bf r}_{12}$ connecting the centers of the black holes,
cf.\ Appendix B. In the following we identify the bare masses of both
solutions.

The post-Newtonian expansions of the functions $\phi^{\rm BL}$ and $\phi^{\rm 
ML}$ can be written as follows
\bea
\label{}
\phi^{\rm BL}&=&\phi^{\rm BL}_{(2)}+\phi^{\rm BL}_{(4)}+\phi^{\rm BL}_{(6)}
+\phi^{\rm BL}_{(8)}+\phi^{\rm BL}_{(10)}
+{\cal O}\left(\frac{1}{c^{12}}\right),
\\[2ex]
\phi^{\rm ML}&=&\phi^{\rm ML}_{(2)}+\phi^{\rm ML}_{(4)}+\phi^{\rm ML}_{(6)}
+\phi^{\rm ML}_{(8)}+\phi^{\rm ML}_{(10)}
+{\cal O}\left(\frac{1}{c^{12}}\right),
\eea
where $\phi_{(n)}$ are functions of $\frac{n}{2}$PN order, as they belong to
the three metric. The details of the expansions are given in Appendices A and
B. The results read
\bea
\label{pne02}
\ds\phi^{\rm BL}_{(2)} &=& \phi^{\rm ML}_{(2)}
\;=\; \frac{1}{4\pi} \left(\frac{m_1}{r_1}+ \frac{m_2}{r_2}\right),
\\[2ex]
\label{pne04}
\ds\phi^{\rm BL}_{(4)} &=& \phi^{\rm ML}_{(4)}
\;=\; -\frac{2}{(16\pi)^2}\frac{m_1m_2}{r_{12}}
\left(\frac{1}{r_1}+ \frac{1}{r_2}\right),
\\[2ex]
\label{pne06}
\ds\phi^{\rm BL}_{(6)} &=& \phi^{\rm ML}_{(6)}
\;=\; \frac{1}{(16\pi)^3}m_1m_2(m_1+m_2)\frac{1}{r_{12}^2}
\left(\frac{1}{r_1}+\frac{1}{r_2}\right),
\\[2ex]
\label{pnb08}
\ds\phi^{\rm BL}_{(8)}
&=& -\frac{1}{2(16\pi)^4}m_1m_2(m_1^2+3m_1m_2+m_2^2)\frac{1}{r_{12}^3}
\left(\frac{1}{r_1}+\frac{1}{r_2}\right),
\\[2ex]
\label{pnb10}
\ds\phi^{\rm BL}_{(10)}
&=&
\frac{1}{4(16\pi)^5}m_1m_2(m_1+m_2)(m_1^2+5m_1m_2+m_2^2)\frac{1}{r_{12}^4}
\left(\frac{1}{r_1}+\frac{1}{r_2}\right),
\\[2ex]
\label{pnm08}
\ds\phi^{\rm ML}_{(8)}
&=&  \phi^{\rm BL}_{(8)} + \frac{1}{2(16\pi)^4}\frac{m_1m_2}{r_{12}^2} \left(
m_2^2\,\frac{{\bf n}_2\cdot{\bf n}_{12}}{r_2^2}
- m_1^2\,\frac{{\bf n}_1\cdot{\bf n}_{12}}{r_1^2} \right),
\\[2ex]
\label{pnm10}
\ds\phi^{\rm ML}_{(10)}
&=& \phi^{\rm BL}_{(10)} + \frac{1}{4(16\pi)^5}\frac{m_1m_2}{r_{12}^3} \left[
m_1^2(m_1+6m_2)\frac{{\bf n}_1\cdot{\bf n}_{12}}{r_1^2}
- m_2^2(m_2+6m_1)\frac{{\bf n}_2\cdot{\bf n}_{12}}{r_2^2}
\right.\nonumber\\[2ex]&&\left.
\ds  - m_1m_2(m_1+m_2)\frac{1}{r_{12}}
\left(\frac{1}{r_1}+\frac{1}{r_2}\right)
\right].
\eea

The equations (\ref{pne02})--(\ref{pnm10}) show that the ML solution at the
4PN order of approximation attributes a dipole moment to each black hole
whereas the BL solution, as it is already evident from the exact expression
(\ref{BL}),  shows monopoles only.  In shifting the centers of the black holes
one can  arrange that also in case of the ML solution the dipole moments do
vanish. To  show this let us introduce
\bea
\phi^{\rm BL}_{\le5{\rm PN}}({\bf x};{\bf x}_1,{\bf x}_2)
&:=&\phi^{\rm BL}_{(2)}+\phi^{\rm BL}_{(4)}
+\phi^{\rm BL}_{(6)}+\phi^{\rm BL}_{(8)}+\phi^{\rm BL}_{(10)},
\\[2ex]
\phi^{\rm ML}_{\le5{\rm PN}}({\bf x};{\bf x}_1,{\bf x}_2)
&:=&\phi^{\rm ML}_{(2)}+\phi^{\rm ML}_{(4)}
+\phi^{\rm ML}_{(6)}+\phi^{\rm ML}_{(8)}+\phi^{\rm ML}_{(10)}.
\eea
Then the shifted ML solution can be defined as
\be
\phi^{\rm ML\,shifted}_{\le5{\rm PN}}({\bf x};{\bf x}_1,{\bf x}_2)
:= \phi^{\rm ML}_{\le5{\rm PN}}({\bf x};
{\bf x}_1+\alpha{\bf r}_{12},{\bf x}_2+\beta{\bf r}_{21}),
\ee
where $\alpha$ and $\beta$ are some dimensionless parameters. We have found that 
for
\bea
\alpha=\frac{1}{(16\pi)^3}\frac{m_1^2 m_2}{8r_{12}^3}
-\frac{1}{(16\pi)^4}\frac{m_1^2(m_1 m_2+5m_2^2)}{16r_{12}^4},
\\[2ex]
\beta=\frac{1}{(16\pi)^3}\frac{m_1 m_2^2}{8r_{12}^3}
-\frac{1}{(16\pi)^4}\frac{m_2^2(m_1 m_2+5m_1^2)}{16r_{12}^4},
\eea
the shifted ML solution coincides with the BL solution up to the 5PN order of 
approximation:
\be
\phi^{\rm ML\,shifted}_{\le5{\rm PN}}({\bf x};{\bf x}_1,{\bf x}_2)
=\phi^{\rm BL}_{\le5{\rm PN}}({\bf x};{\bf x}_1,{\bf x}_2)
+{\cal O}\left(\frac{1}{c^{12}}\right).
\ee

The Hamiltonian we calculate by means of formula
\be
\label{sur}
H = -\lim_{R\rightarrow\infty}
\kern-1ex \oint\limits_{S\left(0,R\right)}
\kern-2ex d\sigma_i\,\phi_{,i},
\ee
where $S\left(0,R\right)$ is a sphere of radius $R$ centered at the origin of 
the coordinate system. Making use of Eqs.\ (\ref{BL}) and (\ref{ML}) we obtain
that the BL and ML solutions lead to the Hamiltonians
\bea
\label{hablt}
H^{\rm BL} &=& 32\pi\left(\alpha_1+\alpha_2\right),
\\[2ex]
\label{hamlt}
H^{\rm ML} &=& 32\pi\left(a+b\right)
+32\pi\sum_{n=2}^\infty\left(a_n+b_n\right).
\eea

Using Eqs.\ (\ref{hablt}) and (\ref{hamlt}) [or Eqs.\
(\ref{pne02})--(\ref{pnm10}) together with Eq.\ (\ref{sur})] we calculate the
static Hamiltonian up to the 3PN order of approximation (notice: the $n$PN
Hamiltonian is determined  by the $(n+2)$PN three metric). We obtain, dropping
the total mass $m_1+m_2$ contribution (in the reduced variables \cite{rv})
\bea
\label{habl}
\widehat{H}^{\rm BL}_{\le3{\rm PN}}=-\lim_{R\rightarrow\infty}
\kern-2ex \oint\limits_{S\left(0,R\right)}
\kern-2ex d\sigma_i \left({\phi^{\rm BL}_{\le5{\rm PN}}}\right)_{,i}
= -\frac{1}{r}+\frac{1}{2r^2}-\frac{1}{4}(1+\nu)\frac{1}{r^3}
+\frac{1}{8}(1+3\nu)\frac{1}{r^4},
\\[2ex]
\label{haml}
\widehat{H}^{\rm ML}_{\le3{\rm PN}}=-\lim_{R\rightarrow\infty}
\kern-2ex \oint\limits_{S\left(0,R\right)}
\kern-2ex d\sigma_i \left({\phi^{\rm ML}_{\le5{\rm PN}}}\right)_{,i}
= -\frac{1}{r}+\frac{1}{2r^2}-\frac{1}{4}(1+\nu)\frac{1}{r^3}
+\frac{1}{8}(1+2\nu)\frac{1}{r^4}.
\eea
Obviously, the both 3PN Hamiltonians are different. The difference  vanishes,
however, if in the case of the ML solution the shifted solution  for the
potential function $\phi$ is used in the calculation of the  Hamiltonian.
Therefore, the postulate of vanishing black-hole dipole moments in isotropic
coordinates yields a unique static 3PN Hamiltonian for binary black holes. The 
condition of vanishing dipole moments needs the metric coefficients; it can not 
be formulated on the Hamiltonian level alone.

\section{Binary point-mass calculations}

In this section we show the results of the static binary point-mass
calculations. In the static case the Hamiltonian constraint equation for the
two-body point-mass system in the canonical formalism of ADM reads
\be
\label{ce1}
\left(1+\frac{1}{8}\phi\right)\Delta\phi=-\sum_a m_a\delta_a
\ee
(for binary systems all sums run over $a=1,2$). Eq.\ (\ref{ce1}) yields the
following formal expansion
\be
\label{ce2}
\Delta\phi=-\left(1+\frac{1}{8}\phi\right)^{-1}\sum_a m_a\delta_a
=-\left(\sum_a m_a\delta_a\right) 
\sum_{n=0}^\infty\left(-\frac{1}{8}\phi\right)^n.
\ee
Using Eq.\ (\ref{ce2}) we obtain the Hamiltonian constraint equations valid at 
individual orders in $1/c$. They read
\bea
\label{lap2}
\Delta\phi_{(2)}&=&-\sum_a m_a\delta_a,
\\[2ex]
\Delta\phi_{(4)}&=&\frac{1}{8}\phi_{(2)}\sum_a m_a\delta_a,
\\[2ex]
\Delta\phi_{(6)}&=&\left(-\frac{1}{64}\phi_{(2)}^2+\frac{1}{8}\phi_{(4)}\right)
\sum_a m_a\delta_a,
\\[2ex]
\Delta\phi_{(8)}&=&\left(\frac{1}{512}\phi_{(2)}^3
-\frac{1}{32}\phi_{(2)}\phi_{(4)}+\frac{1}{8}\phi_{(6)}\right)
\sum_a m_a\delta_a,
\\[2ex]
\label{lap10}
\Delta\phi_{(10)}&=&\left(-\frac{1}{4096}\phi_{(2)}^4
+\frac{3}{512}\phi_{(2)}^2\phi_{(4)}-\frac{1}{64}\phi_{(4)}^2
-\frac{1}{32}\phi_{(2)}\phi_{(6)}+\frac{1}{8}\phi_{(8)}\right)
\sum_a m_a\delta_a.
\eea

All Poisson equations (\ref{lap2})--(\ref{lap10}) are of the form
\be
\label{poi}
\Delta\phi=\sum_a f({\bf x})\delta_a,
\ee
where the function $f$ is usually singular at ${\bf x}={\bf x}_a$. One can 
propose three different ways of solving equations of type (\ref{poi}). The 
first two ways are based on the following sequence of equalities:
\be
\label{reg12}
\phi=\Delta^{-1}\left(\sum_a f({\bf x})\delta_a\right)
= \Delta^{-1}\left(\sum_a f_{\rm reg}({\bf x}_a)\delta_a\right)
= \sum_a f_{\rm reg}({\bf x}_a)\Delta^{-1}\delta_a
= -\frac{1}{4\pi}\sum_a f_{\rm reg}({\bf x}_a)\frac{1}{r_a},
\ee
where $f_{\rm reg}({\bf x}_a)$ is the regularized value of the function $f$ at
${\bf x}={\bf x}_a$, defined by means of the Hadamard's ``partie finie"
procedure \cite{freg}. The difference between the two first methods relies on
different evaluating the regular value of the products of singular functions.
In the first method we use
\be
\label{reg1}
\left(f_1({\bf x})f_2({\bf x})\right) \delta_a
= f_{1\rm reg}({\bf x}_a) f_{2\rm reg}({\bf x}_a) \delta_a,
\ee
whereas in the second method instead of the rule (\ref{reg1}) we apply
\be
\label{reg2}
\left(f_1({\bf x})f_2({\bf x})\right) \delta_a
= \left(f_1f_2\right)_{\rm reg}({\bf x}_a) \delta_a.
\ee
For the developments in the book by Infeld and Pleba{\'n}ski \cite{IP60},  it
was crucial that the both regularization procedures coincided, i.e.,
$(f_1f_2)_{\rm reg}=f_{1\rm reg}f_{2\rm reg}$  (``tweedling of products"). In
the third method we regularize the Poisson integral rather than the source 
function:
\be
\label{reg3}
\phi = -\frac{1}{4\pi} \sum_a \int d^3x'
\frac{f({\bf x}')\delta({\bf x}'-{\bf x}_a)}{|{\bf x}'-{\bf x}|}
= -\frac{1}{4\pi} \sum_a
\left(\frac{f({\bf x}')}{|{\bf x}'-{\bf x}|}\right)_{\rm reg}
({\bf x}'={\bf x}_a).
\ee

Let us denote the results of applying the regularization method based on Eqs.\ 
(\ref{reg12}) and (\ref{reg1}) by primes, the results of the method based on 
Eqs.\ (\ref{reg12}) and (\ref{reg2}) by double primes, and the results of the 
method based on Eq.\ (\ref{reg3}) by triple primes. It turns out that up to 
$1/c^6$ the results of the three methods coincide and, moreover, they are 
identical with the results of the expansions of the BL and ML solutions:
\be
\label{phi1}
\phi'_{(n)}=\phi''_{(n)}=\phi'''_{(n)}=\phi^{\rm BL}_{(n)}=\phi^{\rm ML}_{(n)},
\quad n=2,4,6.
\ee
The results of the first method coincide with the expansion of the BL solution,
what we have checked up to the 5PN order:
\be
\label{phi2}
\phi'_{(n)}=\phi^{\rm BL}_{(n)},\quad n=2,4,6,8,10.
\ee
The function $\phi''_{(8)}$ calculated by means of the second method coincides 
with $\phi'_{(8)}$ (and $\phi^{\rm BL}_{(8)}$), and the function $\phi''_{(10)}$
reads
\be
\phi''_{(10)} = \phi^{\rm BL}_{(10)}
+ \frac{1}{2(16\pi)^5}\frac{m_1^2m_2^2}{r_{12}^4}
\left(\frac{m_1}{r_1}+\frac{m_2}{r_2}\right).
\ee
The functions $\phi'''_{(8)}$ and $\phi'''_{(10)}$ are equal to
\bea
\ds\phi'''_{(8)} &=& \phi^{\rm BL}_{(8)}
+ \frac{1}{2(16\pi)^4}\frac{m_1m_2}{r_{12}^2} \left(
m_1^2\,\frac{{\bf n}_1\cdot{\bf n}_{12}}{r_1^2}
- m_2^2\,\frac{{\bf n}_2\cdot{\bf n}_{12}}{r_2^2} \right),
\\[2ex]
\label{phi5}
\ds\phi'''_{(10)} &=& \phi^{\rm BL}_{(10)}
+ \frac{1}{4(16\pi)^5}\frac{m_1m_2}{r_{12}^3} \left\{
m_2^2(m_2+6m_1) \frac{{\bf n}_2\cdot{\bf n}_{12}}{r_2^2}
- m_1^2(m_1+6m_2) \frac{{\bf n}_1\cdot{\bf n}_{12}}{r_1^2}
\right.\nonumber\\[2ex]&&\left.
\ds + \frac{m_1m_2}{r_{12}} \left[ 
\left(2m_1+m_2\right)\frac{1}{r_1}+\left(2m_2+m_1\right)\frac{1}{r_2} \right]
\right\}.
\eea
Notice, the mass parameters $m_1$ and $m_2$ in Eqs.\ (\ref{phi1})--(\ref{phi5})
denote total rest masses of infinitely separated bodies whereas the mass
parameters in Eqs.\ (\ref{ce1})--(\ref{lap10}) denote some formal rest masses
only; the former results from the latter by our regularization procedures. The
total rest-mass parameters we identify with the bare masses of the BL and ML
solutions.

By means of Eq.\ (\ref{sur}) we calculate the static Hamiltonian up to the 3PN 
order applying the functions $\phi'_{(n)}$, $\phi''_{(n)}$, and
$\phi'''_{(n)}$. The results are (dropping the total mass $m_1+m_2$
contribution):
\bea
\label{h'}
\widehat{H}'^{\,{\rm static}}_{\le3{\rm PN}}=-\lim_{R\rightarrow\infty}
\kern-2ex \oint\limits_{S\left(0,R\right)}
\kern-2ex d\sigma_i \left({\phi'_{\le5{\rm PN}}}\right)_{,i}
= -\frac{1}{r}+\frac{1}{2r^2}-\frac{1}{4}(1+\nu)\frac{1}{r^3}
+\frac{1}{8}(1+3\nu)\frac{1}{r^4},
\\[2ex]
\label{h''}
\widehat{H}''^{\,{\rm static}}_{\le3{\rm PN}}=-\lim_{R\rightarrow\infty}
\kern-2ex \oint\limits_{S\left(0,R\right)}
\kern-2ex d\sigma_i \left({\phi''_{\le5{\rm PN}}}\right)_{,i}
= -\frac{1}{r}+\frac{1}{2r^2}-\frac{1}{4}(1+\nu)\frac{1}{r^3}
+\frac{1}{8}(1+4\nu)\frac{1}{r^4},
\\[2ex]
\label{h'''}
\widehat{H}'''^{\,{\rm static}}_{\le3{\rm PN}}=-\lim_{R\rightarrow\infty}
\kern-2ex \oint\limits_{S\left(0,R\right)}
\kern-2ex d\sigma_i \left({\phi'''_{\le5{\rm PN}}}\right)_{,i}
= -\frac{1}{r}+\frac{1}{2r^2}-\frac{1}{4}(1+\nu)\frac{1}{r^3}
+\frac{1}{8}(1+\frac{9}{2}\nu)\frac{1}{r^4}.
\eea
Obviously, at the 3PN order of approximation, the three Hamiltonians  differ
from each other and from the Hamiltonian (\ref{haml}) obtained from the
expanded ML solution. The Hamiltonian  (\ref{h'}) coincides with the
Hamiltonian (\ref{habl}) obtained using the PN expansion of the BL solution,
what obviously follows from Eq.\ (\ref{phi2}).

\section{Consistency calculations and comparison with the previous results}

In the region $\Omega:=B\left(0,R\right)\setminus 
\left[B\left({\bf x}_1,\ve_1\right)\cup B\left({\bf x}_2,\ve_2\right)\right]$
(where $B\left({\bf x}_a,\ve_a\right)$ ($a=1,2$) is a ball of radius $\ve_a$
around the position ${\bf x}_a$ of the $a$th body and $B\left(0,R\right)$
is a ball of radius $R$ centered at the origin of the coordinate system) the
right-hand sides of the Eqs.\ (\ref{lap2})--(\ref{lap10}) vanish, so the
functions $\phi_{(n)}$ fulfil the Laplace equation in this region. Applying
Gauss's theorem we thus obtain
\be
\label{loc1}
0 = \int\limits_\Omega d^3x\,\Delta\phi_{(n)}
= \kern-2ex \oint\limits_{\pa B\left({\bf x}_1,\ve_1\right)}
\kern-3ex d\sigma_i\,\phi_{(n),i}
+ \kern-2ex \oint\limits_{\pa B\left({\bf x}_2,\ve_2\right)}
\kern-3ex d\sigma_i\,\phi_{(n),i}
+ \kern-2ex \oint\limits_{\pa B\left(0,R\right)}
\kern-2ex d\sigma_i\,\phi_{(n),i},
\ee
with the normal vectors pointing inwards the spheres
$\pa B\left({\bf x}_a,\ve_a\right)$ and outwards the sphere 
$\pa B\left(0,R\right)$. From Eq.\ (\ref{loc1}) it follows that
\be
\label{loc2}
\lim_{R\rightarrow\infty} \kern-2ex \oint\limits_{\pa B\left(0,R\right)}
\kern-2ex d\sigma_i\,\phi_{(n),i}
= -\lim_{\ve_1\rightarrow0}
\kern-2ex \oint\limits_{\pa B\left({\bf x}_1,\ve_1\right)}
\kern-3ex d\sigma_i\,\phi_{(n),i}
- \lim_{\ve_2\rightarrow0}
\kern-2ex \oint\limits_{\pa B\left({\bf x}_2,\ve_2\right)}
\kern-3ex d\sigma_i\,\phi_{(n),i},
\ee
so the Hamiltonian $H_{n{\rm PN}}$ at the $n$PN order, according to Eq.\
(\ref{sur}), can be calulated as
\be
\label{loc3}
H_{n{\rm PN}} =  -\lim_{\ve_1\rightarrow0}
\kern-2ex \oint\limits_{\pa B\left({\bf x}_1,\ve_1\right)}
\kern-3ex d\sigma_i\,\phi_{(2n+4),i}
- \lim_{\ve_2\rightarrow0}
\kern-2ex \oint\limits_{\pa B\left({\bf x}_2,\ve_2\right)}
\kern-3ex d\sigma_i\,\phi_{(2n+4),i}.
\ee
We have used Eq.\ (\ref{loc3}) and the functions $\phi'_{(n)}$, $\phi''_{(n)}$,
and $\phi'''_{(n)}$, to calculate the static Hamiltonian up to the 3PN order.
The integrals over the spheres $\pa B\left({\bf x}_a,\ve_a\right)$ diverge as
$\ve_a\to0$, so to calculate them we have used the Hadamard's procedure
\cite{freg}. The results coincide with those given by Eqs.\
(\ref{h'})--(\ref{h'''}).

The $n$PN Hamiltonian can also be written in the form of a volume
integral
\be
H_{n{\rm PN}} = - \int d^3x\,\Delta\phi_{(2n+4)},
\ee
so still another way of calculating it relies on direct integration of the
(minus) right-hand sides of Eqs.\ (\ref{lap2})--(\ref{lap10}). To do this one
must use the Hadamard's regularization \cite{freg} together with the rule
(\ref{reg1}) or (\ref{reg2}). For the functions $\phi'_{(n)}$ we have used the
rule (\ref{reg1}) (as in deriving the functions $\phi'_{(n)}$), whereas for the
functions $\phi''_{(n)}$ and $\phi'''_{(n)}$ we have applied the rule
(\ref{reg2}). The results coincide again with those given by Eqs.\
(\ref{h'})--(\ref{h'''}).

In the paper \cite{JS98} we applied the `double prime' regularization 
procedure and had thus obtained the Hamiltonian (\ref{h''}); see Sec.\ VI in
Ref.\ \cite{JS98}. The same result we had obtained from the static $n$-body
Hamiltonian of Ref.\ \cite{KT72} by applying some expansion-and-limiting
procedure. If we take the $n$-body static Hamiltonian of \cite{KT72} but
specialize it to the two-body case adapting only those terms which are directly
finite, we get the BL result of Eq.\ (\ref{habl}). It corresponds to our
`single prime' regularization procedure described above.

\bigskip\noindent
{\bf Acknowledgment}

\noindent The authors thank Thibault Damour for useful comments. 
This work was supported by the Max-Planck-Gesellschaft Grant No.\ 
02160-361-TG74 (GS).

\appendix

\section{Brill-Lindquist solution}

We use the representation of the Brill-Lindquist (BL) solution taken from
Appendix A of \cite{AP96}. The BL solution can be written in the form [cf.\
Eq.\ (3) in \cite{AP96}]
\be
\label{a1}
\phi^{\rm BL} = 8\left(\frac{\alpha_1}{r_1}+\frac{\alpha_2}{r_2}\right).
\ee
The bare masses $m_1$ and $m_2$ of the black holes depend on the parameters
$\alpha_1$, $\alpha_2$ and the coordinate distance $r_{12}$ between the black
holes [cf.\ Eqs.\ (A7) and (A8) in \cite{AP96}]:
\be
\label{a2}
m_1 = 32\pi\alpha_1\left(1+\frac{\alpha_2}{r_{12}}\right),\quad
m_2 = 32\pi\alpha_2\left(1+\frac{\alpha_1}{r_{12}}\right).
\ee
The unique positive solutions of Eqs.\ (\ref{a2}) for $\alpha_1$ and $\alpha_2$
read
\bea
\label{a3a}
\alpha_1 &=& -\frac{1}{4}\left(2r_{12}+\frac{m_2-m_1}{16\pi}\right)
+ \frac{1}{4}r_{12}
\sqrt{4+\frac{m_1+m_2}{4\pi r_{12}}
+\left(\frac{m_1-m_2}{16\pi r_{12}}\right)^2},
\\[2ex]
\label{a3b}
\alpha_2 &=& -\frac{1}{4}\left(2r_{12}+\frac{m_1-m_2}{16\pi}\right)
+ \frac{1}{4}r_{12}
\sqrt{4+\frac{m_1+m_2}{4\pi r_{12}}
+\left(\frac{m_1-m_2}{16\pi r_{12}}\right)^2}.
\eea

We expand the right-hand sides of Eqs.\ (\ref{a3a}) and (\ref{a3b}) in powers
of $1/c$ taking into account that the masses $m_1$ and $m_2$ can be regarded 
as being of order $1/c^2$. The results we substitute into Eq.\ (\ref{a1}) to
obtain the post-Newtonian expansion of the function $\phi^{\rm BL}$. Such
obtained functions $\phi^{\rm BL}_{(n)}$ for $n=2,4,6,8,10$ are given in Eqs.\
(\ref{pne02}), (\ref{pne04}), (\ref{pne06}), (\ref{pnb08}), and (\ref{pnb10}),
respectively.

\section{Misner-Lindquist solution}

The form of the Misner-Lindquist (ML) solution we use is taken from Appendix B
of \cite{AP96}. The ML solution we write in the form [cf.\ Eq.\ (4) in
\cite{AP96}]
\be
\label{b1}
\phi^{\rm ML} = 8\left(\frac{a}{r_1}+\frac{b}{r_2}\right)
 + 8\sum_{n=2}^{\infty}
\left(\frac{a_n}{|{\bf x}-{\bf d}_n|}+\frac{b_n}{|{\bf x}-{\bf e}_n|} \right),
\ee
where $r_a=|{\bf x}-{\bf x}_a|$ ($a=1,2$) and ${\bf x}_a$ is the position of
the center of the $a$th black hole relative to a given origin in the flat
space, $a$ and $b$ are the radii of the black hole 1 and 2, respectively; ${\bf
d}_n$ ($n\ge2$) are the positions of the image poles of black hole 1, ${\bf
e}_n$ ($n\ge2$) are the positions of the image poles of black hole 2, $a_n$ and
$b_n$ ($n\ge2$) are the corresponding weights.

The black hole 1 together with its odd images and the even images of the black
hole 2 are located on the positive $z$ axis, whereas the black hole 2 together
with its odd images and the even images of the black hole 1 lie on the negative
$z$ axis. The relative distances $|{\bf x}-{\bf d}_n|$ and $|{\bf x}-{\bf
e}_n|$ entering Eq.\ (\ref{b1}) can be expressed by radii $a$ and $b$ of the
black holes and their relative position vector ${\bf r}_{12}$ as follows (here
$n\ge2$)
\be
\label{b2}
\begin{array}{rlll}
|{\bf x}-{\bf d}_n|
&=& \sqrt{r_1^2 + D_n^2 + 2 r_1 D_n ({\bf n}_1\cdot{\bf n}_{12})}\quad
&\mbox{for $n$ odd},
\\[2ex]
|{\bf x}-{\bf d}_n|
&=& \sqrt{r_2^2 + D'^2_n - 2 r_2 D'_n ({\bf n}_2\cdot{\bf n}_{12})}\quad
&\mbox{for $n$ even},
\\[2ex]
|{\bf x}-{\bf e}_n|
&=& \sqrt{r_2^2 + E'^2_n - 2 r_2 E'_n ({\bf n}_2\cdot{\bf n}_{12})}\quad
&\mbox{for $n$ odd},
\\[2ex]
|{\bf x}-{\bf e}_n|
&=& \sqrt{r_1^2 + E_n^2 + 2 r_1 E_n ({\bf n}_1\cdot{\bf n}_{12})}\quad
&\mbox{for $n$ even},
\end{array}
\ee 
where
\be
\label{b3}
\begin{array}{rlll}
D_n &=& \ds r_{12}\left(1-\frac{\sinh[(n+1)\mu_0]}
{\sinh[(n+1)\mu_0]+\frac{a}{b}\sinh[(n-1)\mu_0]}\right)
\quad &\mbox{for $n$ odd},
\\[3ex]
D'_n &=& \ds r_{12}\left(1-\frac{ab\sinh[(n+2)\mu_0]+a^2\sinh n\mu_0}
{r_{12}^2\sinh n\mu_0}\right)
\quad &\mbox{for $n$ even},
\\[3ex]
E_n &=& \ds r_{12}\left(1-\frac{ab\sinh[(n+2)\mu_0]+b^2\sinh n\mu_0}
{r_{12}^2\sinh n\mu_0}\right)
\quad &\mbox{for $n$ even},
\\[3ex]
E'_n &=& \ds r_{12}\left(1-\frac{\sinh[(n+1)\mu_0]}
{\sinh[(n+1)\mu_0]+\frac{b}{a}\sinh[(n-1)\mu_0]}\right)
\quad &\mbox{for $n$ odd}.
\end{array}
\ee
The quantity $\mu_0$ entering Eqs.\ (\ref{b3}) is given by
\be
\label{b4}
\cosh2\mu_0 = \frac{r_{12}^2-a^2-b^2}{2ab}.
\ee
The weights $a_n$ and $b_n$ ($n\ge2$) of the image poles depend on the radii
$a$ and $b$ of the black holes and the distance $r_{12}$ between them:
\be
\label{b5}
\begin{array}{rlll}
a_n &=& \ds\frac{ab\sinh2\mu_0}{r_{12}\sinh n\mu_0}
&\mbox{for $n$ even},
\\[2ex]
a_n &=& \ds\frac{ab\sinh2\mu_0}{b\sinh[(n+1)\mu_0]+a\sinh[(n-1)\mu_0]}
&\mbox{for $n$ odd},
\\[2ex]
b_n &=& \ds\frac{ab\sinh2\mu_0}{r_{12}\sinh n\mu_0}
&\mbox{for $n$ even},
\\[2ex]
b_n &=& \ds\frac{ab\sinh2\mu_0}{a\sinh[(n+1)\mu_0]+b\sinh[(n-1)\mu_0]}
&\mbox{for $n$ odd}.
\end{array}
\ee

To obtain the post-Newtonian expansion of the ML solution we expand the
right-hand side of Eq.\ (\ref{b1}) in powers of $1/c$ taking into account that
the radii $a$ and $b$ are of order $1/c^2$. To do this we use Eqs.\
(\ref{b2})--(\ref{b5}). We have found that only the first five terms from the
sum on the right-hand side of Eq.\ (\ref{b1}) contributes to the 5PN order. The
result of the expansion reads
\bea
\label{b6}
\frac{1}{8}\phi^{\rm ML} &=& \frac{a}{r_1}+\frac{b}{r_2}
+ \frac{ab}{r_{12}}\left(\frac{1}{r_1}+\frac{1}{r_2}\right)
+ \frac{ab}{r_{12}^2}\left(\frac{a}{r_1}+\frac{b}{r_2}\right)
\nonumber\\[2ex]&&
+ \frac{a^2b^2}{r_{12}^3}\left(\frac{1}{r_1}+\frac{1}{r_2}\right)
+ \frac{ab}{r_{12}^2}
\left(\frac{b^2({\bf n}_2\cdot{\bf n}_{12})}{r_2^2}
-\frac{a^2({\bf n}_1\cdot{\bf n}_{12})}{r_1^2}\right)
\nonumber\\[2ex]&&
+ \frac{a^2b^2(a+b)}{r_{12}^4}\left(\frac{1}{r_1}+\frac{1}{r_2}\right)
+ \frac{ab}{r_{12}^3}
\left(\frac{b^3({\bf n}_2\cdot{\bf n}_{12})}{r_2^2}
-\frac{a^3({\bf n}_1\cdot{\bf n}_{12})}{r_1^2}\right)
+ {\cal O}\left(\frac{1}{c^{12}}\right).
\eea

For the ML solution we use the definition of the bare masses $m_1$ and $m_2$
introduced by Lindquist in \cite{L63}.  The masses $m_1$ and $m_2$ depend on
the radii $a$ and $b$ of the black holes and the distance $r_{12}$ between the
centers of the black holes [cf.\ Eqs.\ (B19) and (B20)]:
\bea
\label{b7a}
m_1 &=& \frac{32\pi ab}{r_{12}}\sinh2\mu_0 \sum_{n=1}^\infty n \left\{
\frac{2}{\sinh2n\mu_0}
+ \frac{1}{\sinh(2n\mu_0-\mu_2)}
+ \frac{1}{\sinh(2n\mu_0+\mu_2)} \right\},
\\[2ex]
\label{b7b}
m_2 &=& \frac{32\pi ab}{r_{12}}\sinh2\mu_0 \sum_{n=1}^\infty n \left\{
\frac{2}{\sinh2n\mu_0}
+ \frac{1}{\sinh[2(n+1)\mu_0-\mu_2]}
+ \frac{1}{\sinh[2(n-1)\mu_0-\mu_2]} \right\},\quad
\eea
where $\mu_2$ is given by
\be
\sinh \mu_2 = \frac{a}{r_{12}}\sinh2\mu_0.
\ee
We have iteratively solved Eqs.\ (\ref{b7a}) and (\ref{b7b}) with respect to
$a$ and $b$. The result, valid up to the 5PN order of approximation, read
\bea
\label{b8}
a &=& \frac{1}{16\pi} \frac{m_1}{2}
- \frac{1}{(16\pi)^2} \frac{m_1m_2}{2r_{12}}
+ \frac{1}{(16\pi)^3} \frac{m_1m_2(2m_1+3m_2)}{8r_{12}^2}
- \frac{1}{(16\pi)^4} \frac{m_1m_2(m_1^2+4m_1m_2+2m_2^2)}{8r_{12}^3}
\nonumber\\[2ex]&&
+ \frac{1}{(16\pi)^5}
\frac{m_1m_2(2m_1^3+14m_1^2m_2+18m_1m_2^2+5m_2^3)}{32r_{12}^4}
+{\cal O}\left(\frac{1}{c^{12}}\right),
\eea
the equation for $b$ can be obtained from the above one by replacements
$m_1\to m_2$ and $m_2\to m_1$.

To obtain the PN expansion of the ML solution we substitute Eqs.\ (\ref{b8})
into Eq.\ (\ref{b6}). Such obtained functions $\phi^{\rm ML}_{(n)}$ for
$n=2,4,6,8,10$ are given in Eqs.\ (\ref{pne02}), (\ref{pne04}), (\ref{pne06}),
(\ref{pnm08}), and (\ref{pnm10}), respectively.


\begin{thebibliography}{99}
\bibitem{rv} The reduced relative position vector ${\bf r}$ is defined as ${\bf
r}:=16\pi\left({\bf x}_1-{\bf x}_2\right)/M$, together with $r:=|{\bf r}|$ and
${\bf n}:={\bf r}/r$; here $M=m_1+m_2$. We also introduce the parameter
$\nu:=\mu/M$, where $\mu:=m_1 m_2/M$ is the reduced mass of the system.  As the
reduced Hamilton function $\widehat{H}$ we define $\widehat{H}:=H/\mu$.
$\widehat{H}$ depends on masses of the binary system only through the parameter
$\nu$. From now on the hat will indicate division by the reduced mass $\mu$.
\end{thebibliography}
\end{document}